\pdfoutput=1

\documentclass[10pt]{article}
\usepackage{graphicx}

\def\Title#1{\begin{center} {\Large #1 } \end{center}}
\def\Author#1{\begin{center}{ \sc #1} \end{center}}
\def\Address#1{\begin{center}{ \it #1} \end{center}}

\newcommand\pubblock{\rightline{\begin{tabular}{l} Proceedings of the Fifth Annual LHCP\\ \pubnumber\\
         \pubdate  \end{tabular}}}

\newenvironment{Abstract}{\begin{quotation} \begin{center} 
             \large ABSTRACT \end{center}\bigskip 
      \begin{center}\begin{large}}{\end{large}\end{center} \end{quotation}}

\newenvironment{Presented}{\begin{quotation} \begin{center} 
             PRESENTED AT\end{center}\bigskip 
      \begin{center}\begin{large}}{\end{large}\end{center} \end{quotation}}

\usepackage{ifthen} 
\newboolean{uprightparticles}
\setboolean{uprightparticles}{false} 
\usepackage{xspace} 
\usepackage{upgreek}

\def\lhcb {\mbox{LHCb}\xspace}

\def\herschel {\mbox{\textsc{HeRSCheL}}\xspace}
\def\MagUp {\mbox{\em Mag\kern -0.05em Up}\xspace}

\ifthenelse{\boolean{uprightparticles}}%
{

 \def\Pmu         {\ensuremath{\upmu}\xspace}

 \def\Ppsi        {\ensuremath{\uppsi}\xspace}

 \def\PDelta      {\ensuremath{\Delta}\xspace}                 
 \def\PXi      {\ensuremath{\Xi}\xspace}                 
 \def\PLambda      {\ensuremath{\Lambda}\xspace}                 
 \def\PSigma      {\ensuremath{\Sigma}\xspace}                 
 \def\POmega      {\ensuremath{\Omega}\xspace}                 
 \def\PUpsilon      {\ensuremath{\Upsilon}\xspace}                 
 
 \def\PB      {\ensuremath{\mathrm{B}}\xspace}                 
                  
 \def\PD      {\ensuremath{\mathrm{D}}\xspace}

 \def\PJ      {\ensuremath{\mathrm{J}}\xspace}                 
 \def\PK      {\ensuremath{\mathrm{K}}\xspace}

 \def\PW      {\ensuremath{\mathrm{W}}\xspace}

 \def\PZ      {\ensuremath{\mathrm{Z}}\xspace}                 
                  
 \def\Pb      {\ensuremath{\mathrm{b}}\xspace}

 \def\Pi      {\ensuremath{\mathrm{i}}\xspace}

}
{

 \def\Pmu         {\ensuremath{\mu}\xspace}

 \def\Ppsi        {\ensuremath{\psi}\xspace}                 
                  
 \mathchardef\PDelta="7101
 \mathchardef\PXi="7104
 \mathchardef\PLambda="7103
 \mathchardef\PSigma="7106
 \mathchardef\POmega="710A
 \mathchardef\PUpsilon="7107
                  
 \def\PB      {\ensuremath{B}\xspace}                 
                  
 \def\PD      {\ensuremath{D}\xspace}

 \def\PJ      {\ensuremath{J}\xspace}                 
 \def\PK      {\ensuremath{K}\xspace}

 \def\PW      {\ensuremath{W}\xspace}

 \def\PZ      {\ensuremath{Z}\xspace}                 
                  
 \def\Pb      {\ensuremath{b}\xspace}

 \def\Pi      {\ensuremath{i}\xspace}

}

\makeatletter
\ifcase \@ptsize \relax
  \newcommand{\miniscule}{\@setfontsize\miniscule{4}{5}}
\or
  \newcommand{\miniscule}{\@setfontsize\miniscule{5}{6}}
\or
  \newcommand{\miniscule}{\@setfontsize\miniscule{5}{6}}
\fi
\makeatother

\DeclareRobustCommand{\optbar}[1]{\shortstack{{\miniscule (\rule[.5ex]{1.25em}{.18mm})}
  \\ [-.7ex] $#1$}}

\def\mup        {{\ensuremath{\Pmu^+}}\xspace}
\def\mun        {{\ensuremath{\Pmu^-}}\xspace} 

\def\Wp     {{\ensuremath{\PW^+}}\xspace}
\def\Wm     {{\ensuremath{\PW^-}}\xspace}

\def\Z      {{\ensuremath{\PZ}}\xspace}

\def\bquark    {{\ensuremath{\Pb}}\xspace}
\def\bquarkbar {{\ensuremath{\overline \bquark}}\xspace}
\def\bbbar     {{\ensuremath{\bquark\bquarkbar}}\xspace}

  \def\Kbar    {{\kern 0.2em\overline{\kern -0.2em \PK}{}}\xspace}

\def\KorKbar    {\kern 0.18em\optbar{\kern -0.18em K}{}\xspace}

  \def\Dbar    {{\kern 0.2em\overline{\kern -0.2em \PD}{}}\xspace}

\def\DorDbar    {\kern 0.18em\optbar{\kern -0.18em D}{}\xspace}

\def\Bbar    {{\ensuremath{\kern 0.18em\overline{\kern -0.18em \PB}{}}}\xspace}

\def\BorBbar    {\kern 0.18em\optbar{\kern -0.18em B}{}\xspace}

\def\jpsi     {{\ensuremath{{\PJ\mskip -3mu/\mskip -2mu\Ppsi\mskip 2mu}}}\xspace}
\def\psitwos  {{\ensuremath{\Ppsi{(2S)}}}\xspace}

  \def\Y#1S{\ensuremath{\PUpsilon{(#1S)}}\xspace}

\def\Lbar        {{\ensuremath{\kern 0.1em\overline{\kern -0.1em\PLambda}}}\xspace}
\def\LorLbar    {\kern 0.18em\optbar{\kern -0.18em \PLambda}{}\xspace}

\def\BF         {{\ensuremath{\mathcal{B}}}\xspace}

\def\BR         {\BF}

\def\to                 {\ensuremath{\rightarrow}\xspace}

\def\AT#1     {\ensuremath{A_{\mathrm{T}}^{#1}}\xspace}           

\def\C#1      {\ensuremath{\mathcal{C}_{#1}}\xspace}                       
\def\Cp#1     {\ensuremath{\mathcal{C}_{#1}^{'}}\xspace}                    
\def\Ceff#1   {\ensuremath{\mathcal{C}_{#1}^{\mathrm{(eff)}}}\xspace}        
\def\Cpeff#1  {\ensuremath{\mathcal{C}_{#1}^{'\mathrm{(eff)}}}\xspace}       
\def\Ope#1    {\ensuremath{\mathcal{O}_{#1}}\xspace}                       
\def\Opep#1   {\ensuremath{\mathcal{O}_{#1}^{'}}\xspace}                    

\newcommand{\tev}{\ensuremath{\mathrm{\,Te\kern -0.1em V}}\xspace}
\newcommand{\gev}{\ensuremath{\mathrm{\,Ge\kern -0.1em V}}\xspace}
\newcommand{\mev}{\ensuremath{\mathrm{\,Me\kern -0.1em V}}\xspace}
\newcommand{\kev}{\ensuremath{\mathrm{\,ke\kern -0.1em V}}\xspace}
\newcommand{\ev}{\ensuremath{\mathrm{\,e\kern -0.1em V}}\xspace}
\newcommand{\gevc}{\ensuremath{{\mathrm{\,Ge\kern -0.1em V\!/}c}}\xspace}
\newcommand{\mevc}{\ensuremath{{\mathrm{\,Me\kern -0.1em V\!/}c}}\xspace}
\newcommand{\gevcc}{\ensuremath{{\mathrm{\,Ge\kern -0.1em V\!/}c^2}}\xspace}
\newcommand{\gevgevcccc}{\ensuremath{{\mathrm{\,Ge\kern -0.1em V^2\!/}c^4}}\xspace}
\newcommand{\mevcc}{\ensuremath{{\mathrm{\,Me\kern -0.1em V\!/}c^2}}\xspace}

\def\mbarn{\ensuremath{\mathrm{ \,mb}}\xspace}

\def\nb {\ensuremath{\mathrm{ \,nb}}\xspace}

\def\invpb {\ensuremath{\mbox{\,pb}^{-1}}\xspace}

\newcommand{\stat}{\ensuremath{\mathrm{\,(stat)}}\xspace}
\newcommand{\syst}{\ensuremath{\mathrm{\,(syst)}}\xspace}

\def\gsim{{~\raise.15em\hbox{$>$}\kern-.85em
          \lower.35em\hbox{$\sim$}~}\xspace}
\def\lsim{{~\raise.15em\hbox{$<$}\kern-.85em
          \lower.35em\hbox{$\sim$}~}\xspace}

\def\sqs   {\ensuremath{\protect\sqrt{s}}\xspace}

\def\pt         {\mbox{$p_{\mathrm{ T}}$}\xspace}

\def\tell1  {TELL1\xspace}
\def\ukl1   {UKL1\xspace}

\newcommand{\eg}{\mbox{\itshape e.g.}\xspace}

\newcommand{\xx}{\ensuremath{\kern 0.5em }}

\usepackage{multirow} 
\usepackage{booktabs} 
\usepackage{rotating}

\usepackage{multirow} 
\usepackage{rotating} 
\usepackage{dashrule} 
\usepackage{color} 
\usepackage{colortbl} 
\usepackage{graphpap}
\usepackage{xfrac}
\usepackage{amsmath}
\usepackage{lineno}

\newcommand\pubnumber{ LHCb-PROC-2017-030 }

\newcommand\pubdate{August 31, 2017}

\def\affiliation{
On behalf of the LHCb Experiment, \\
Department of Engineering Physics \\
Tsinghua University, Beijing, 100049, China}

\begin{document}

\large
\begin{titlepage}
\pubblock

\vfill
\Title{Measurements of soft QCD and double parton scattering at LHCb}
\vfill

\Author{ Liupan An  }
\Address{\affiliation}
\vfill
\begin{Abstract}
Soft QCD and double parton scattering are of great interest in high energy physics.
They are both actively studied at the LHCb experiment.
The measurement of the central exclusive production of $\jpsi$ and $\psitwos$ mesons in $pp$ collisions at $\sqs = 13\tev$ is presented.
The result shows good agreement with the theoretical predictions.
The measurement of the $\jpsi$ pair production cross-section in $pp$ collisions at $\sqs = 13\tev$ is reported.
The differential cross-sections as functions of various kinematic variables are compared to the theoretical predictions,
and show significant evidence of double parton scattering contribution.

\end{Abstract}
\vfill

\begin{Presented}
The Fifth Annual Conference\\
 on Large Hadron Collider Physics \\
Shanghai Jiao Tong University, Shanghai, China\\ 
May 15-20, 2017
\end{Presented}
\vfill
\end{titlepage}
\def\thefootnote{\fnsymbol{footnote}}
\setcounter{footnote}{0}

\normalsize 

\section{Introduction}
In high energy hadron collisions, 
all the QCD processes can be categorized into the soft and hard ones.
Soft interactions dominate the high energy $pp$ collisions.
They also present in the remains of hard scattering events.
It is essential to model these soft QCD processes in the Monte-Carlo generators of the hard processes of interest.
However, the perturbative QCD breaks down for soft QCD.
Phenomenological models, which use experimental results as input, are necessary to describe it.

A wide range of topics are covered by soft QCD,
\eg properties of the minimum bias data, 
kinematics of the underlying events,
the inelastic cross-section, and the central exclusive production~(CEP).
LHCb has been actively studying each of the topics.
The recent measurement of the central exclusive production of $\jpsi$ and $\psitwos$ mesons in $pp$ collisions at $\sqs = 13\tev$~\cite{CEPpaper} is reported in Sec.~\ref{sec:CEP}. 

\begin{figure}[htb]
\centering
\includegraphics[height=2in]{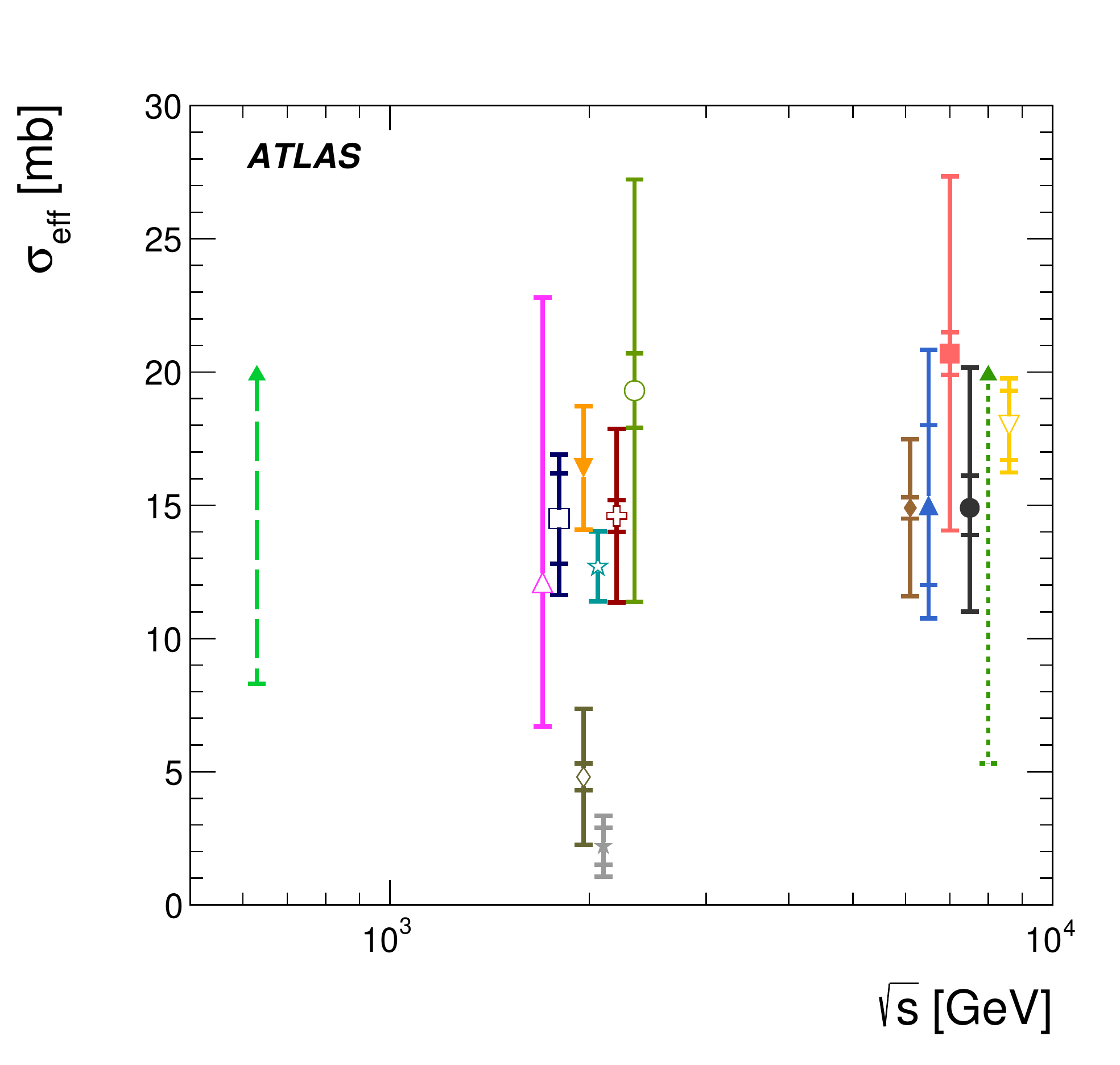}
\includegraphics[height=2in]{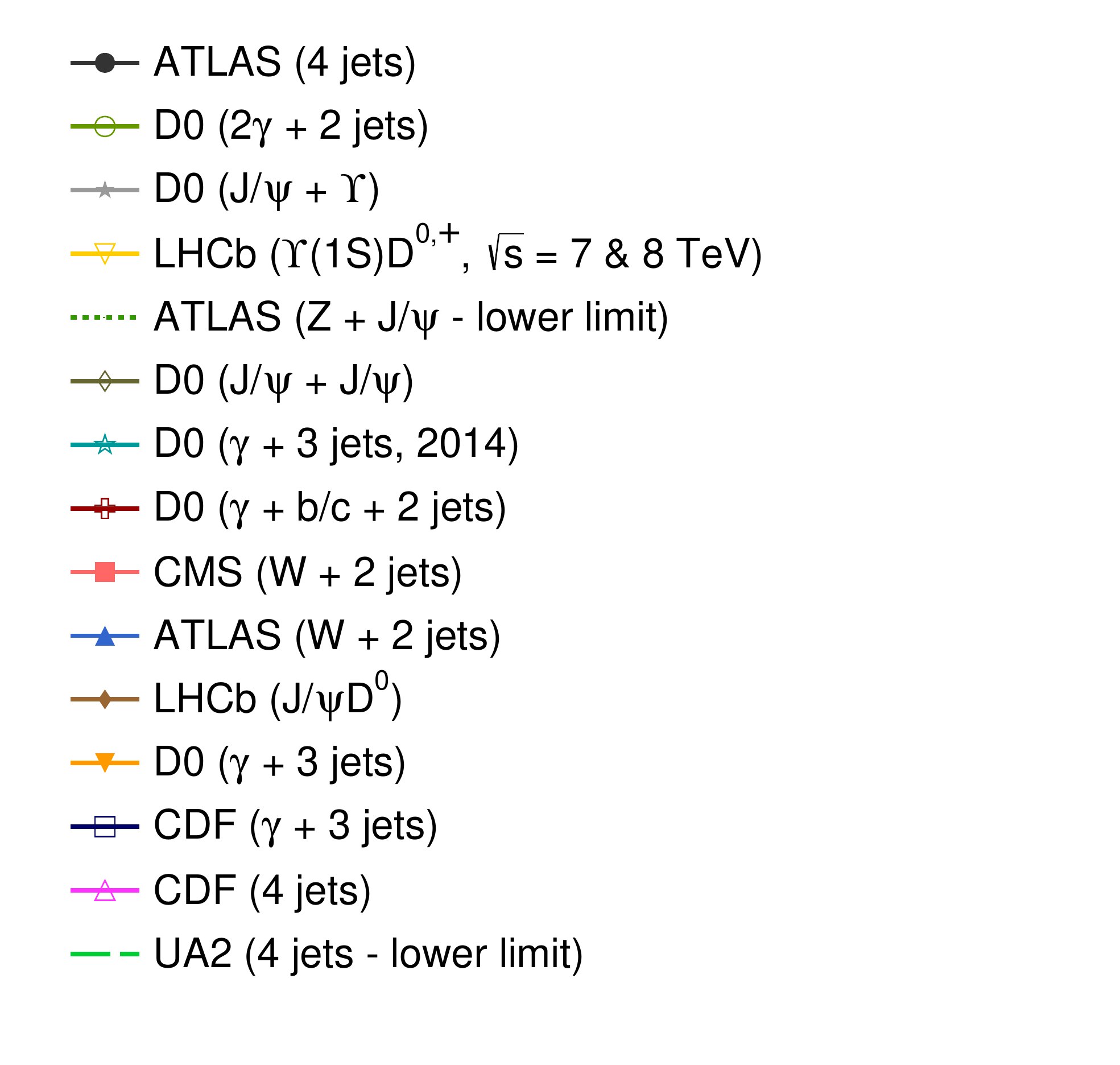}
\caption{Summary of $\sigma_{\mathrm{eff}}$ as a function of the centre-of-mass energy.}
\label{fig:DPS}
\end{figure}

In the hard processes, there can be different numbers of partons participating the scattering.
The process can be categorized as single parton scattering~(SPS), double parton scattering~(DPS) and so on.
The DPS process is of great importance since it can provide information on the parton transverse profile and the parton correlations in proton.
It can also help us to better understand the backgrounds, \eg $\Z+\bbbar$ and $\Wp+\Wm$, in searches for new physics.
Under the assumptions that 
the transverse and longitudinal components of partons can be factorized,
and there is no correlation between the two partons participating the interaction,
the DPS production cross-section is written as
\begin{equation}\label{eq:DPS}
  \sigma_{Q_{1}Q_{2}}^{\mathrm{DPS}} = 
  \dfrac{1}{1 + \delta_{Q_{1}Q_{2}}} 
  \dfrac{ \sigma_{Q_{1}} \sigma_{Q_{2}}}
  {\sigma_{\mathrm{eff}}},
\end{equation}
where $Q_1$ and $Q_2$ indicate the two products,
$\sigma_{Q_{1}}$ and $\sigma_{Q_{2}}$ are their inclusive production cross-sections,
$\delta_{Q_{1}Q_{2}}$ is the symmetry factor,
and $\sigma_{\mathrm{eff}}$ is the effective cross-section.
With these assumptions,
$\sigma_{\mathrm{eff}}$ is thought to be universal for all processes and energies.
DPS measurements can be used to validate these modelling assumptions, 
probing any dependence on process and energy.
A summary of the measured $\sigma_{\mathrm{eff}}$ from various experiments as a function of the centre-of-mass energy is shown in Fig.~\ref{fig:DPS}~\cite{DPSsum}.
Besides, it is determined to be $\sigma_{\mathrm{eff}}=8.2\pm2.2\mbarn$ using $\jpsi$ pair production from the CMS data at $\sqs = 7\tev$~\cite{CMS1,CMS2}. 
The majority of the measurements are consistent and around $15\mbarn$.
But the $\jpsi$ pair production measurements at ATLAS, D0 and CMS, together with the D0 $\jpsi + \PUpsilon$ measurement, 
indicate significant DPS contributions, which leads to smaller $\sigma_{\mathrm{eff}}$ values.
At LHCb, the DPS process has been studied in $\PUpsilon + D$ and $\jpsi + D$.
Evident DPS contributions are observed.
The measured $\sigma_{\mathrm{eff}}$ are around $14.5\mbarn$ and consistent with the majority.
The measurement of the $\jpsi$ pair production cross-section in $pp$ collisions at $\sqs = 13\tev$~\cite{DPSpaper} is reported in Sec.~\ref{sec:DPS}.

\section{CEP at LHCb}
In CEP, which can be denoted as $p+p \to p+X+p$,
there is a fusion of two photons or two pomerons~(a colourless strongly-coupled object), or a photon with a pomeron, 
while the two protons remain intact.
The final state is very clean.
The product $X$ is well isolated in rapidity.

\begin{figure}[htb]
\centering
\includegraphics[height=1.8in]{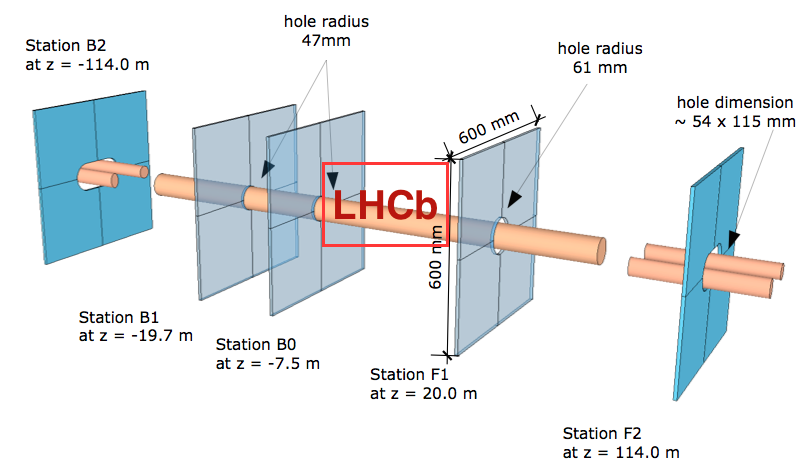}
\caption{Display of the \herschel subdetector.}
\label{fig:herschel}
\end{figure}

The \lhcb detector is a single-arm forward
spectrometer covering the \mbox{pseudorapidity} range $2<\eta <5$.
There are several features making LHCb a major actor in CEP searches.
First, the pile-up level at LHCb is low. 
The CEP processes are much less contaminated.
Second, the vertex locator~(VELO) can provide a backward coverage of $-3.5<\eta <-1.5$,
helping to identify background events with redundant activities.
Third, dedicated CEP trigger lines are developed at LHCb.
They have loose transverse momentum criteria, and require low event multiplicity.
Moreover, in the RunII phase, a new shower counters subdetector, 
\herschel~\cite{hershcel} is implemented, as shown in Fig.~\ref{fig:herschel}.
It consists of five stations of scintillators installed along the beam pipe.
Together with VELO, it extends the \mbox{pseudorapidity} coverage up to $-10<\eta <-5$, $-3.5<\eta <-1.5$ and $1.5<\eta <10$.
The backgrounds in CEP which are most difficult to distinguish from signals are the non-exclusive events,
in which additional gluon emission takes place or one of the protons dissociates,
and the remnants escape the acceptance of the LHCb spectrometer.
\herschel largely enhances the ability to reduce them. 

\section{CEP of $\jpsi$ and $\psitwos$ at $\sqs = 13\tev$}
\label{sec:CEP}
\begin{figure}[htb]
\centering
\includegraphics[height=1.8in]{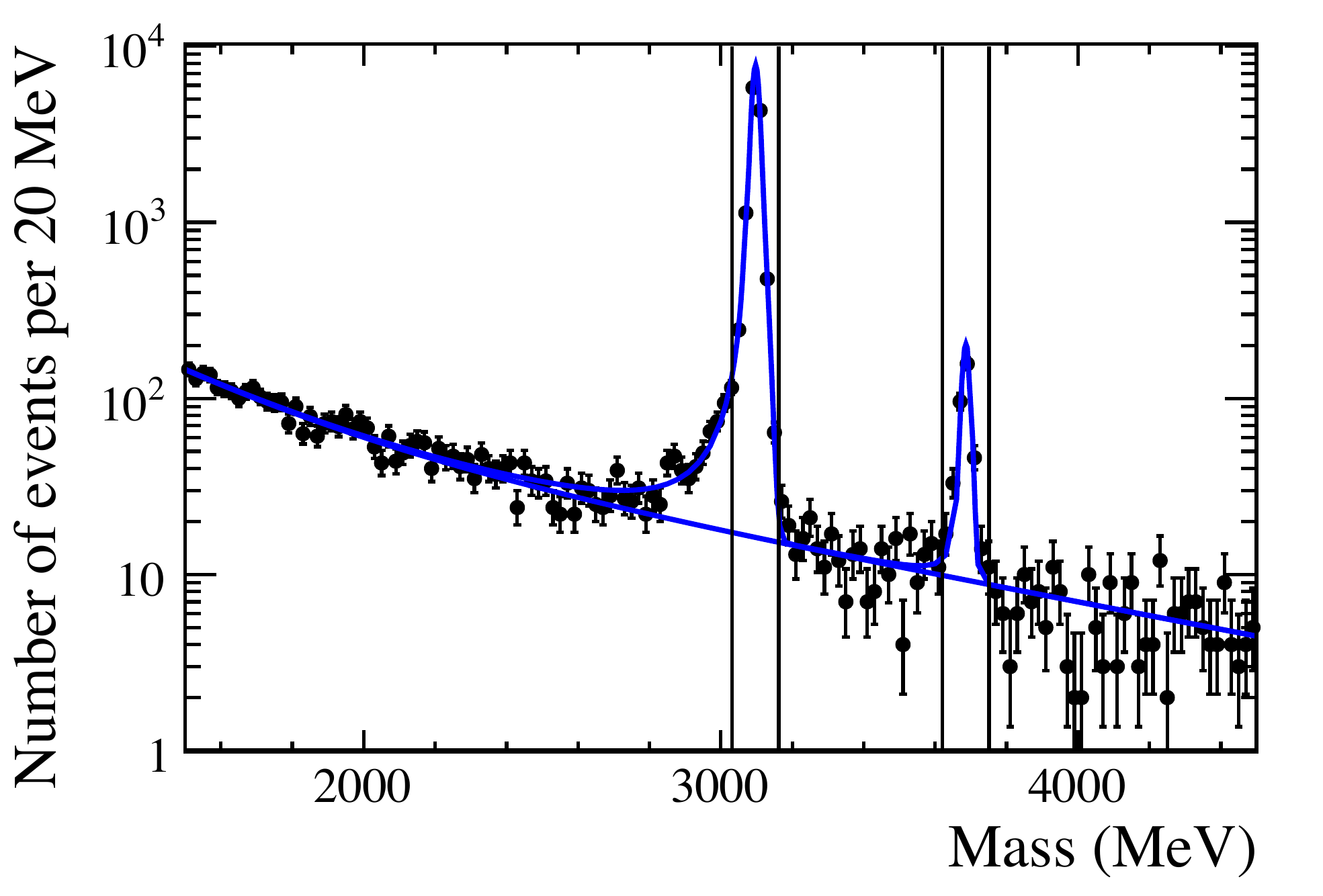}
\caption{Mass distribution of the dimuon candidates after cuts.}
\label{fig:cutmass}
\end{figure}

The CEP cross-sections of $\jpsi$ and $\psitwos$ at $\sqs = 13\tev$ are measured using $204 \invpb$ data collected by LHCb.
Events with additional VELO tracks, 
with reconstructed photons having transverse energy larger than $200 \mevcc$ 
or with significant deposits in \herschel are removed.
Clear mass peaks of $\jpsi$ and $\psitwos$ are obtained after these cuts, as shown in Fig.~\ref{fig:cutmass}.
The inelastic background events are distinguished by performing fit to the squared transverse momentum distribution using two exponential components.
The usage of \herschel roughly halves the background level.

\begin{figure}[htb]
\centering
\includegraphics[height=2in]{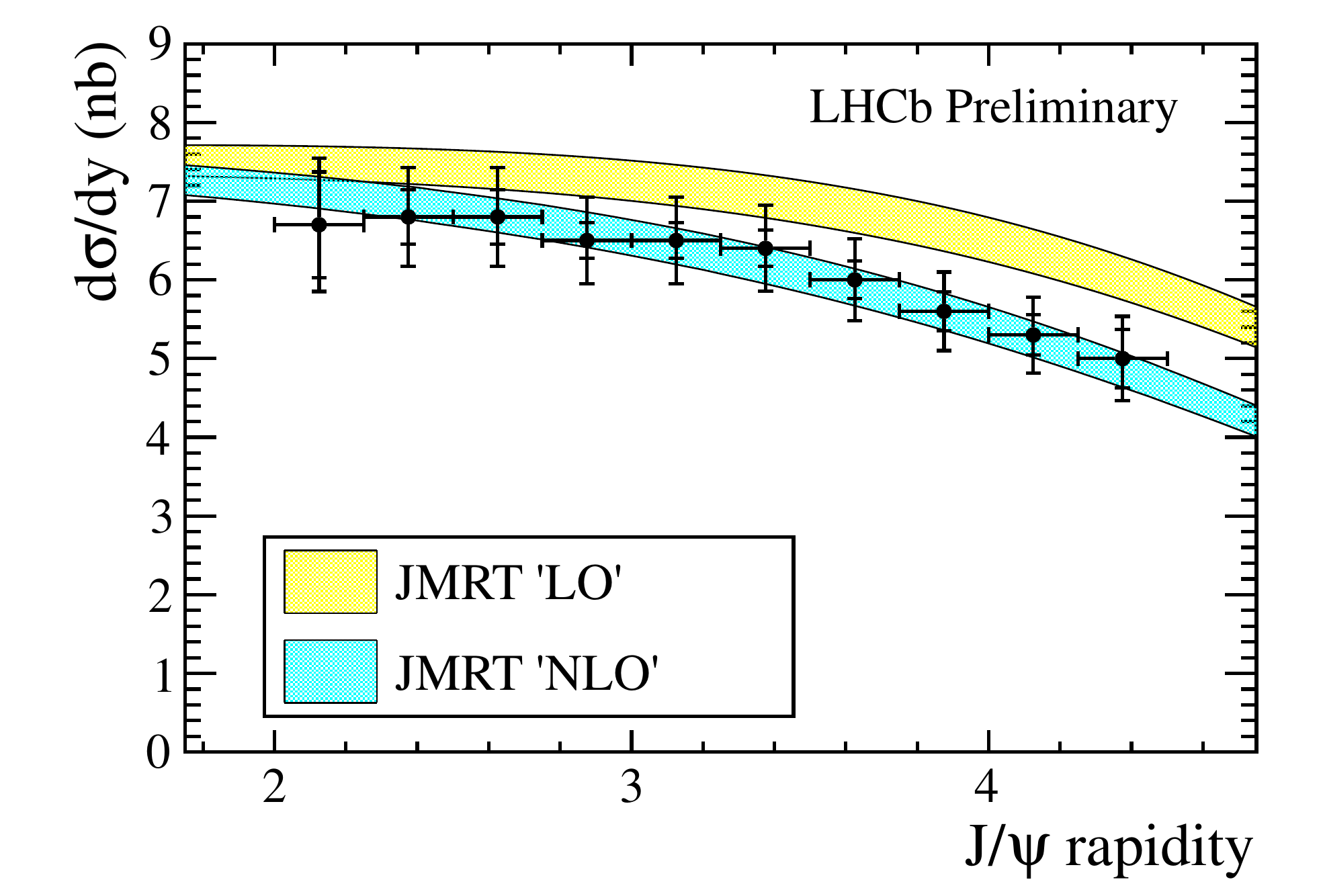}
\includegraphics[height=2in]{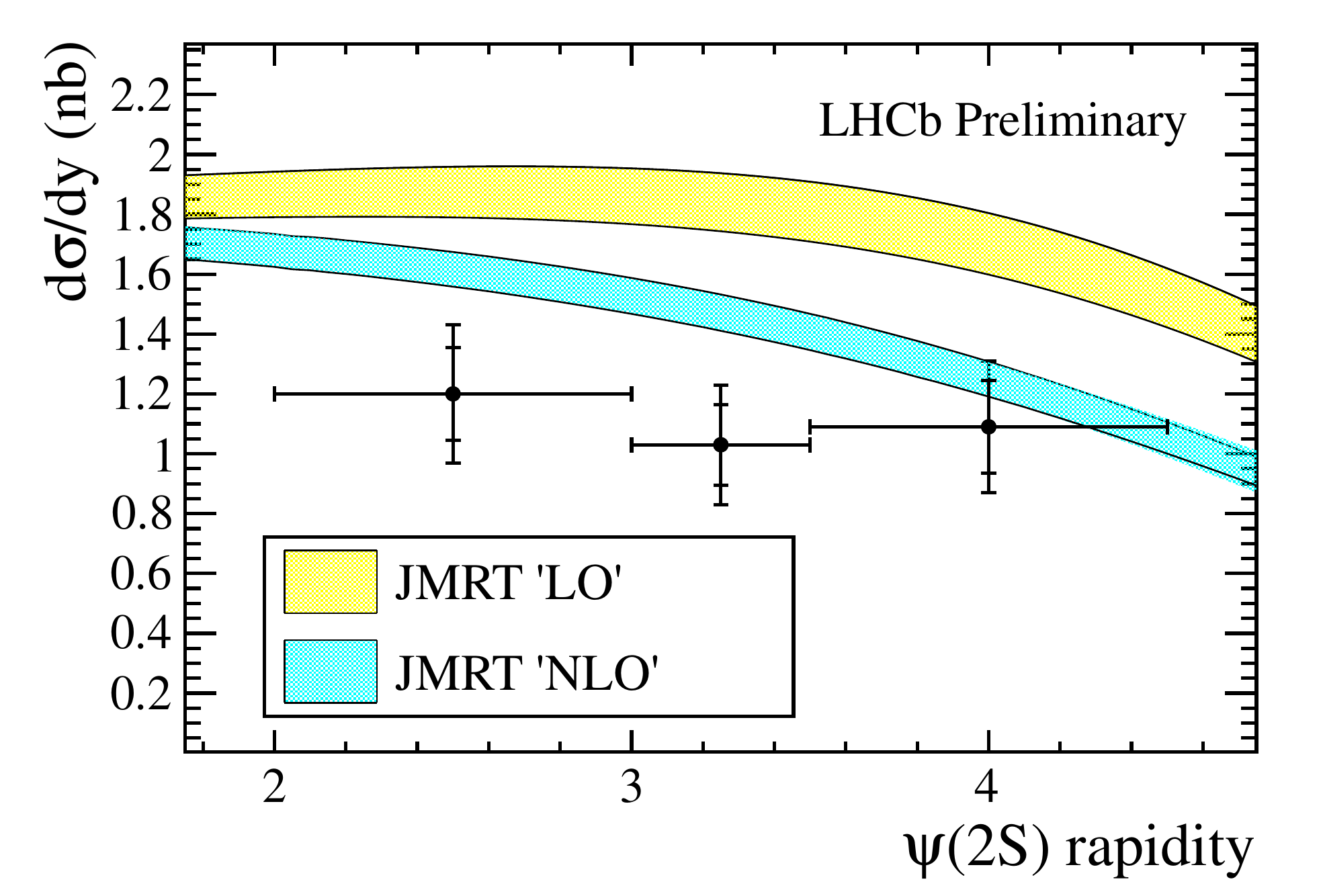}
\caption{Differential cross-sections as a function of the dimuon rapidity for (left)~$\jpsi$ and (right)~$\psitwos$. 
The inner error bars are the statistical uncertainties; the outer bars are the total uncertainties.
The JMRT theoretical predictions are overlaid.}
\label{fig:cmptheory}
\end{figure}

The total cross-sections of exclusive $\jpsi$ and $\psitwos$ to dimuon, 
with both muons having $2<\eta<5$, 
are determined to be 
$$
\begin{array}{rcl}
\sigma_{J/\psi\rightarrow\mu^+\mu^-}(2.0<\eta_{\mu^+},\eta_{\mu^-}<4.5)&=&407\pm 8\pm 24 \pm 16 {\rm \ pb} \\
\sigma_{\psi(2S)\rightarrow\mu^+\mu^-}(2.0<\eta_{\mu^+},\eta_{\mu^-}<4.5)&=&9.4\pm 0.9\pm 0.6 \pm 0.4 {\rm \ pb.}\\
\end{array}
$$
The differential cross-sections as a function of the dimuon rapidity are shown in Fig.~\ref{fig:cmptheory}.
They are compared to the theoretical predictions of the JMRT QCD model at leading order~(LO) and next-to-leading order~(NLO)~\cite{JMRT1,JMRT2}.
The $\jpsi$ cross-section agrees better with the NLO predictions.
The $\psitwos$ cross-section has larger uncertainties,
but also agrees better with the NLO calculations.

\begin{figure}[htb]
\centering
\includegraphics[height=2in]{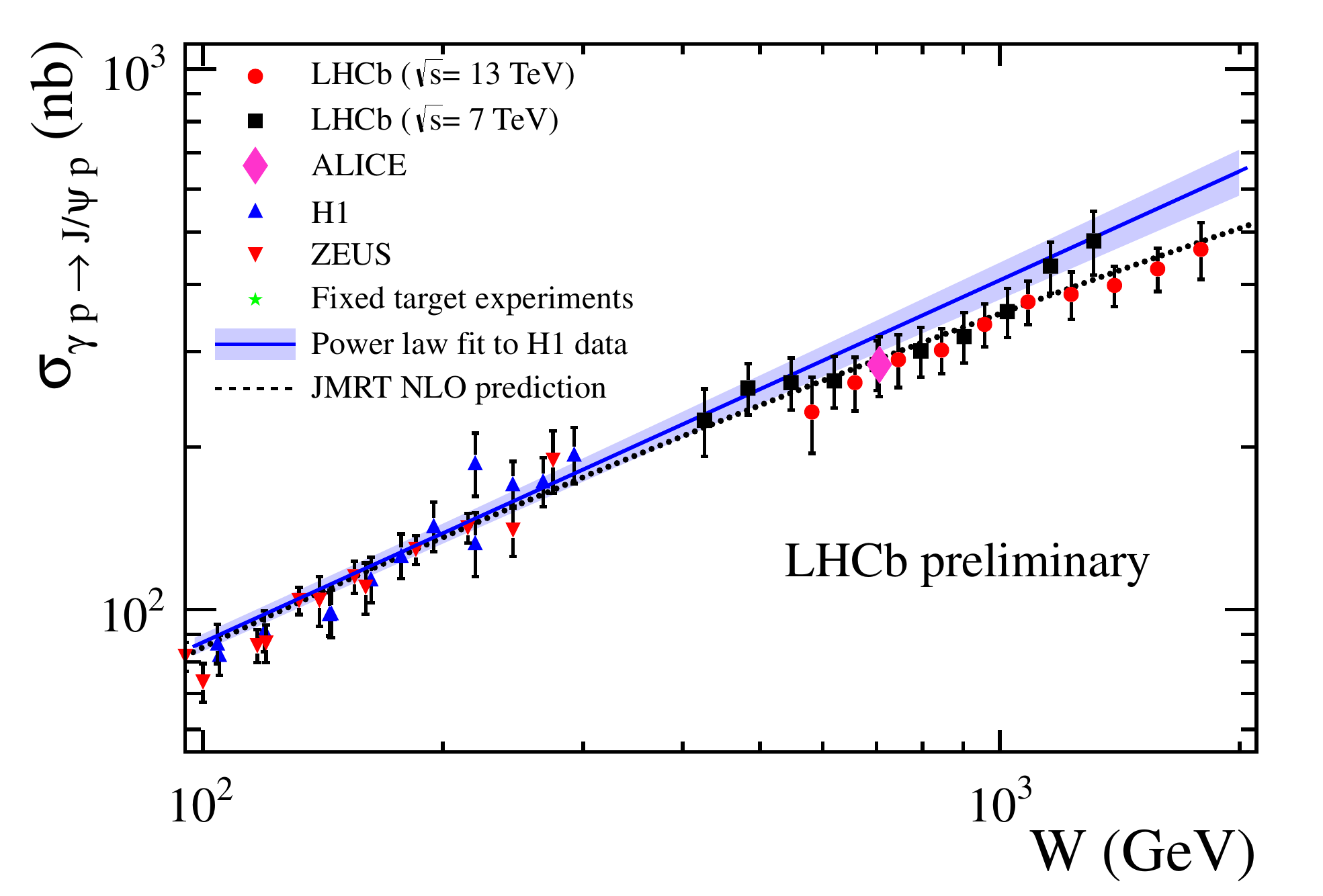}
\includegraphics[height=2in]{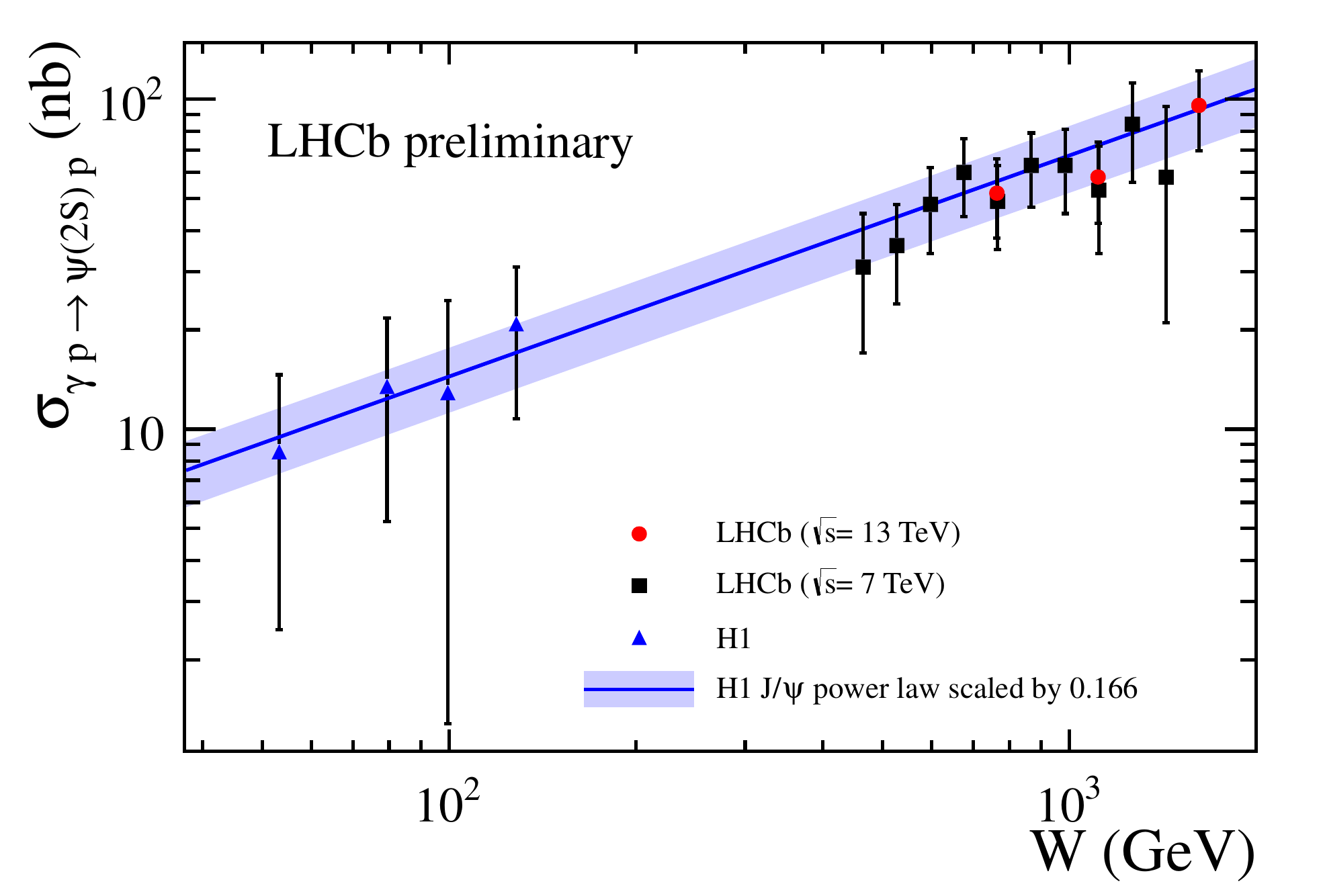}
\caption{Comparison of $\sigma_{\gamma p\rightarrow \psi p}$ as a function of $W$ between different experiments and theoretical predictions for (left)~$\jpsi$ and (right)~$\psitwos$ respectively.
The various components are indicated in the legend.}
\label{fig:photo}
\end{figure}

The CEP cross-section in $pp$ collisions, $\sigma_{pp\rightarrow p\psi p}$, is related to the photo-production cross-section, $\sigma_{\gamma p\rightarrow \psi p}$, via~\cite{JMRT1}
\begin{equation}
\sigma_{pp\rightarrow p\psi p} = 
r(W_+ )k_+{dn\over dk_+} \sigma_{\gamma p\rightarrow \psi p} (W_+)   +
r(W_-) k_-{dn\over dk_-} \sigma_{\gamma p\rightarrow \psi p} (W_-),
\label{eq:photo}
\end{equation}
where $r(W_\pm )$ is the gap survival factor which can be taken from previous studies;
$k_\pm=m_\psi/2 e^{\pm y}$ is the photon energy;
$dn/dk_\pm$ is the photon flux which is also determined in previous studies;
$W_\pm^2=2k_\pm\sqrt{s}$ is the centre-of-mass energy of the photon-proton system.
The $13\tev$ $pp$ collision data collected at LHCb can reach the $W = 2\tev$ photon-proton energy scale, which is the highest so far.
The $\sigma_{\gamma p\rightarrow \psi p}$ as a function of $W$ is compared to the results of other experiments, as shown in Fig.~\ref{fig:photo}.
They are also compared to various theoretical predictions.
The results from different experiments are consistent in the regions where they overlap.
For $\jpsi$, a deviation from the power law fit to the H1 data~\cite{H1} is observed at the highest energies,
while the JMRT NLO prediction~\cite{JMRT1,JMRT2} describes the data well.
For $\psitwos$, the agreement is good for both methods.

\section{$\jpsi$ pair production at $\sqs = 13\tev$}
\label{sec:DPS}
The $\jpsi$ pair production cross-section in $pp$ collisions at $\sqs = 13\tev$ is measured using $279\invpb$ data collected by LHCb
with both $\jpsi$ mesons in the rapidity range $2.0<y<4.5$, and with a transverse momentum~$\pt< 10 \gevc$.
The $\jpsi$ pair production cross-section is measured according to
\begin{equation}\label{eq:crosssection}
  \sigma(\jpsi\jpsi) = \frac{N^{{\rm cor}}}{{\cal{L}} \times \BF(\jpsi \to \mup \mun)^{2}},
\end{equation}
where $N^{{\rm cor}}$ is the signal yield after the per-event efficiency correction,
${\cal{L}}$ is the integrated luminosity,
and \mbox{$\BR(\jpsi \to \mup \mun)$} is the branching fraction of $\jpsi \to \mup \mun$.
The efficiencies are estimated using both simulation and real data.

During the trigger stage, high quality muons are selected.
After that, the identified muons are required to have good track qualities.
They are required to be in the range $\pt>650\mevc$, $6<p<200\gevc$ and $2<\eta<5$.
The two muons coming from one $\jpsi$ meson are required to form a good quality vertex.
The four muons are required to come from the same primary vertex.
Duplicate tracks and multiple candidates are removed.
In the following, the labels $1$ and $2$ are randomly assigned to the two $\jpsi$ candidates.

\begin{figure}[htb]
\centering
\includegraphics[height=2in]{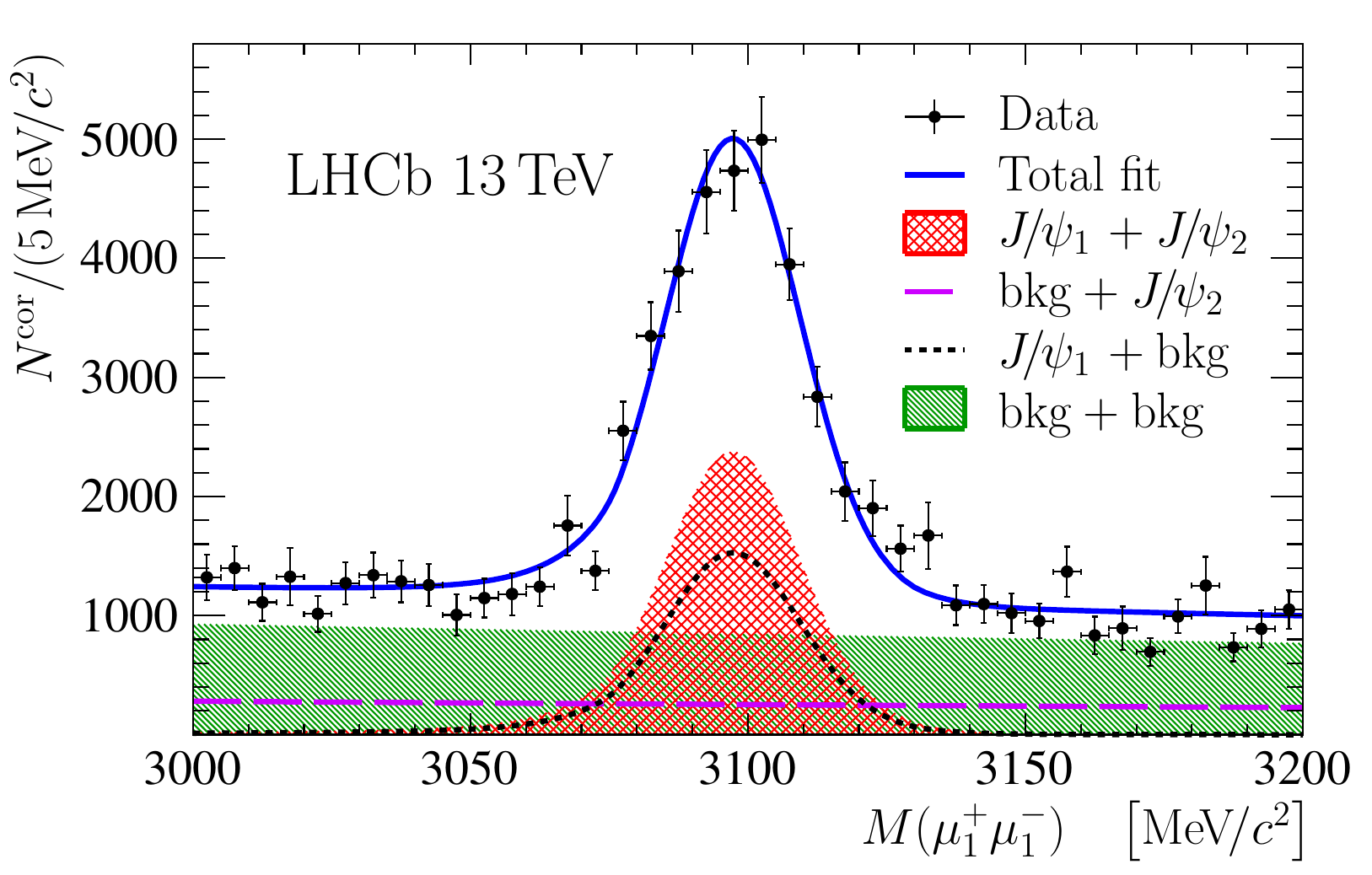}
\includegraphics[height=2in]{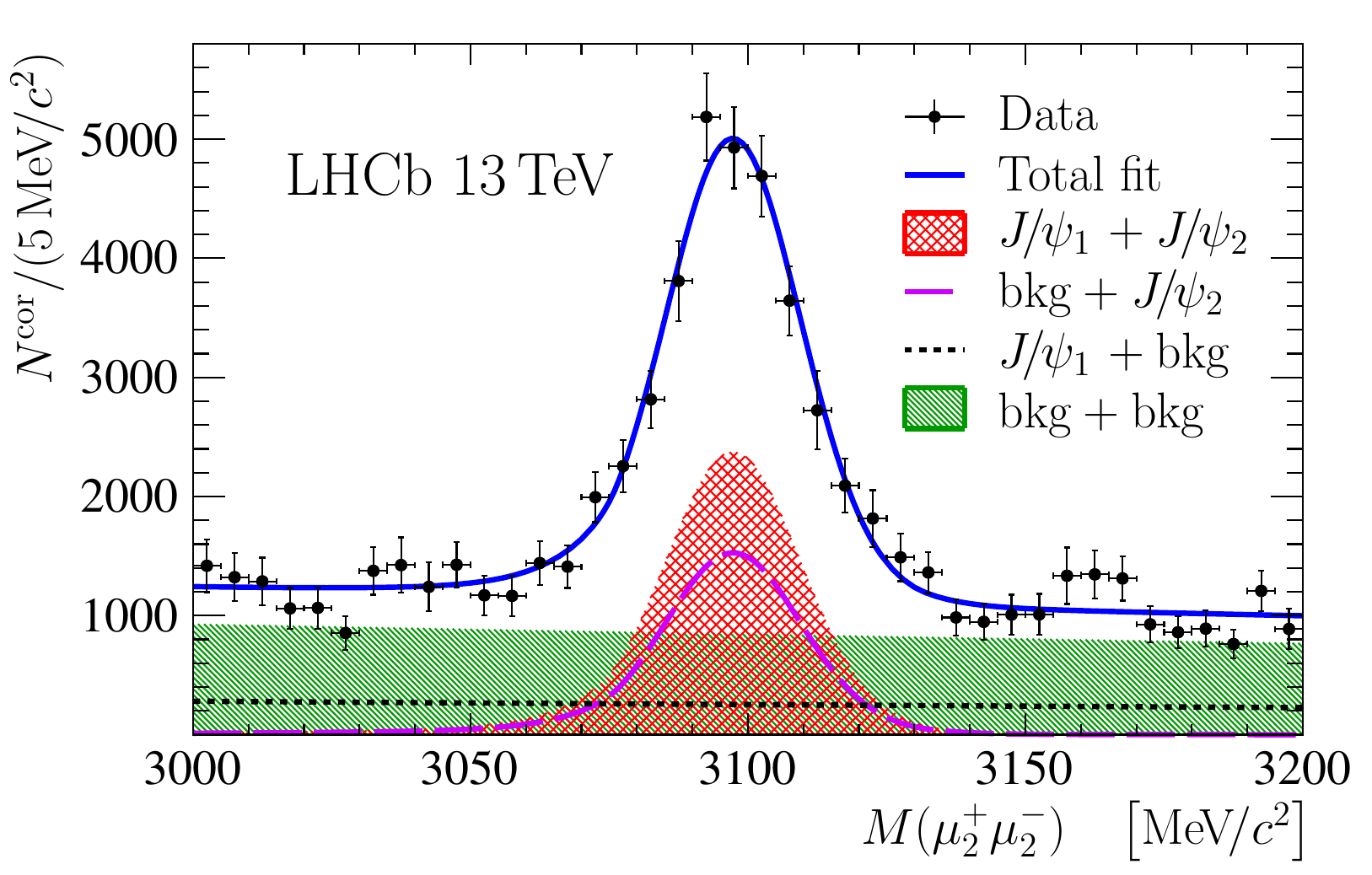}
\caption{
Projections of the fit result on $(M(\mu^+_1\mu^-_1)$ and $M(\mu^+_2\mu^-_2))$.
The various components are indicated in the legend.
}
\label{fig:Jpsifit}
\end{figure}

The signal yield $N^{{\rm cor}}$ is obtained by performing an extended unbinned maximum likelihood fit to 
the efficiency-corrected two-dimensional $(M(\mu^+_1\mu^-_1),M(\mu^+_2\mu^-_2))$ mass distribution.
For each dimension, the signal mass distribution is described by a Gaussian kernel with power tails;
the background is modelled by an exponential function.
The projections of the fit result on $(M(\mu^+_1\mu^-_1)$ and $M(\mu^+_2\mu^-_2))$ are shown in Fig.~\ref{fig:Jpsifit}.
The contribution of the residual contamination of $\jpsi$ from $\bquark$-hadron decays is determined using simulation with
$\sigma(pp \to \bbbar)$ and $\sigma({\rm prompt} \jpsi)$ as input.
The $\jpsi$ pair production cross-section with both $\jpsi$ mesons having \mbox{$2.0<y<4.5$} and \mbox{$\pt < 10 \gevc$} is measured to be
\begin{equation*}
   \sigma(\jpsi\jpsi) = 15.2 \pm 1.0\stat \pm 0.9\syst \nb.
\end{equation*}

\begin{figure}[htb]
\centering
\includegraphics[height=2in]{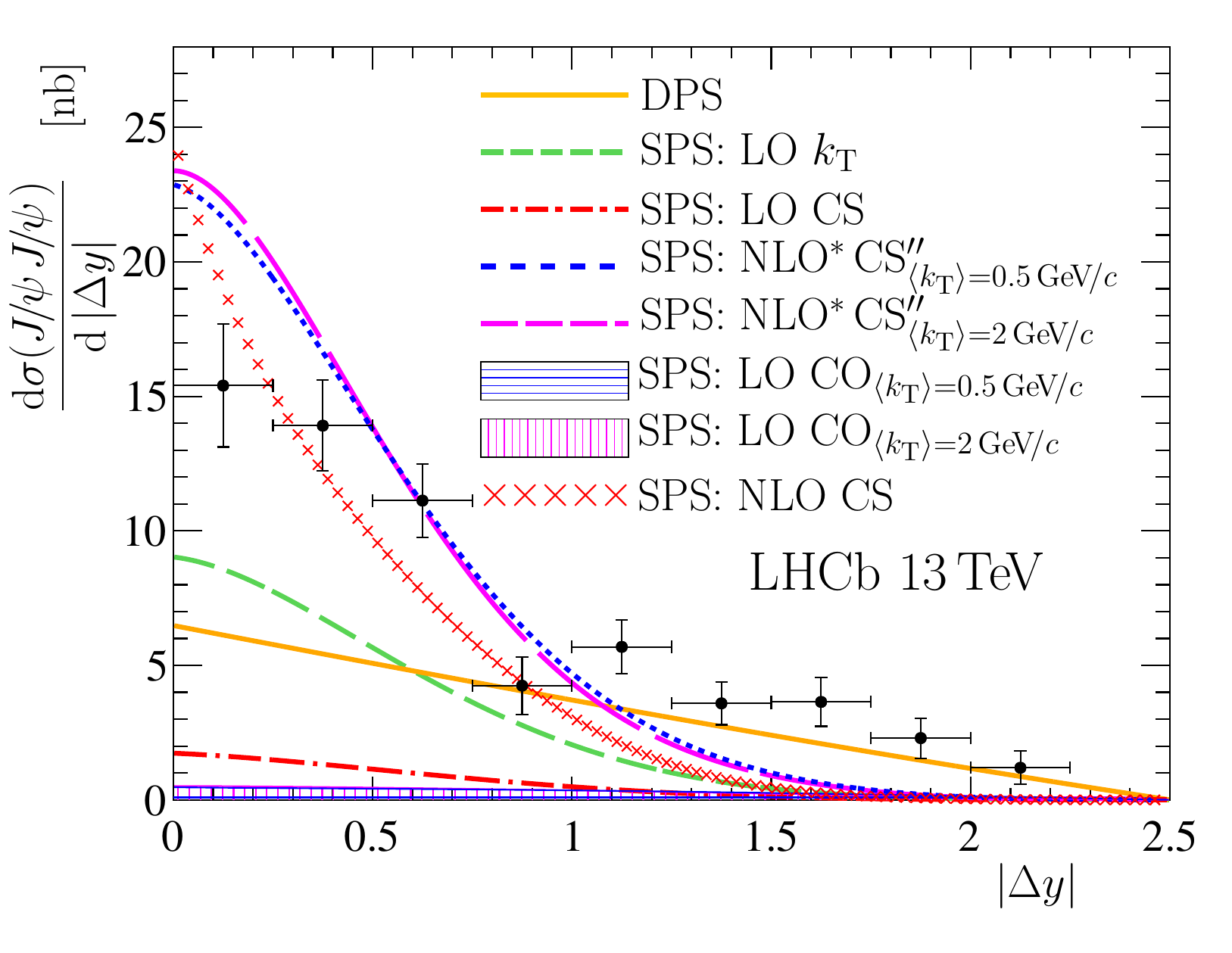}
\caption{
Comparisons between the measurement and various theoretical predictions for the differential cross-section as a function of $\left|\Delta y\right|$.
}
\label{fig:deltay}
\end{figure}

The differential cross-sections of $\jpsi$ pair production with respect to several kinematic variables are compared to various theoretical predictions.
The most significant indication of the DPS contribution comes from the $\left|\Delta y\right|$ distribution, as shown in Fig.~\ref{fig:deltay}.
For $\left|\Delta y\right|>1.5$, there is basically no SPS contribution according to all the theoretical predictions. 
The DPS contribution is essential.
To derive the DPS fraction, 
these distributions are fit with templates that fix the predicted DPS and SPS shapes, with the relative fractions left unconstrained. 
Using the DPS cross-sections obtained from the fit, $\sigma_{\mathrm{eff}}$ is determined according to Eq.~\ref{eq:DPS}.
They are consistent for different models and variables, and lie between $10.0$ and $12.5\mbarn$.
These $\sigma_{\mathrm{eff}}$ values are smaller than the previous LHCb measurements of $\PUpsilon + D$ and $\jpsi + D$. 
They are larger than the ATLAS, D0 and CMS measurements which are also obtained from $\jpsi$ pair production.

\section{Summary}
Soft QCD and DPS are both actively explored at LHCb.
The measurement of the central exclusive production of $\jpsi$ and $\psitwos$ mesons in $pp$ collisions at $\sqs = 13\tev$ is presented.
The result shows good agreement with the JMRT NLO predictions.
The measurement of the $\jpsi$ pair production cross-section in $pp$ collisions at $\sqs = 13\tev$ is also reported.
The differential cross-sections show significant evidence for the DPS contribution.
$\sigma_{\mathrm{eff}}$ is evaluated by performing SPS plus DPS template fits.

\end{document}